\documentclass
[superscriptaddress,secnumarabic,amssymb,amsmath,nobibnotes,aps,prd,showkeys,showpacs,nofootinbib,onecolumn,notitlepage,12pt]{revtex4}%
\usepackage{setspace}
\usepackage{color}
\usepackage{amsmath}
\usepackage{amsfonts}
\usepackage{verbatim}
\usepackage{amssymb}
\usepackage{graphicx,bm}
\usepackage{amsmath}
\usepackage{amssymb}
\usepackage{graphicx}
\usepackage[caption=false]{subfig}%
\providecommand{\U}[1]{\protect\rule{.1in}{.1in}}
\newcommand{\be}{\begin{equation}}
\newcommand{\ee}{\end{equation}}

\newcommand{\mincir}{\raise
-3.truept\hbox{\rlap{\hbox{$\sim$}}\raise4.truept\hbox{$<$}\ }}
\newcommand{\magcir}{\raise
-3.truept\hbox{\rlap{\hbox{$\sim$}}\raise4.truept\hbox{$>$}\ }}

\ifx\pdfoutput\relax\let\pdfoutput=\undefined\fi
\newcount\msipdfoutput
\ifx\pdfoutput\undefined\else
\ifcase\pdfoutput\else
\ifx\paperwidth\undefined\else
\ifdim\paperheight=0pt\relax\else\pdfpageheight\paperheight\fi
\ifdim\paperwidth=0pt\relax\else\pdfpagewidth\paperwidth\fi
\fi\fi\fi
\begin{document}
\title{Cosmological Solutions in Scalar-Tensor theory via the Eisenhart-Duval lift}
\author{Andronikos Paliathanasis}
\email{anpaliat@phys.uoa.gr}
\affiliation{Institute of Systems Science, Durban University of Technology, Durban 4000,
South Africa}
\affiliation{Centre for Space Research, North-West University, Potchefstroom 2520, South Africa}
\affiliation{Departamento de Matem\'{a}ticas, Universidad Cat\'{o}lica del Norte, Avda.
Angamos 0610, Casilla 1280 Antofagasta, Chile}

\begin{abstract}
We implement the Einsenhart-Duval lift in scalar-tensor gravity as a means to
construct integrable cosmological models and analytic cosmological solutions.
Specifically, we employ a geometric criterion to constrain the free functions
of the scalar-tensor theory such that the field equations can be written in
the equivalent form of linear equations. This geometric linearization is
achieved by the introduction of an extended minisuperspace description. The
results are applied to construct analytic solutions in modified theories of
gravity such as the $f\left(  R\right)  $-theory and the hybrid
metric-Palatini $f\left(  \mathcal{R}\right)  $-gravity.

\end{abstract}
\maketitle
\date{\today}

\section{Introduction}

\label{sec1}

Scalar fields play an important role in the description of cosmological
observations \cite{Teg,Kowal,Komatsu,suzuki}. The dynamical degrees of freedom
described by the scalar fields can drive the dynamics to explain the dynamical
behavior of the physical variables, consequently affecting the cosmological
evolution and history \cite{cl1,cl2,cl3,ten1,ten2,ten3,ten4}. There is a
plethora of scalar field models which have been proposed in the literature,
some of which include quintessence \cite{ratra,peebles}, phantom \cite{ph1},
k-essence \cite{kes}, Galileon, quintom \cite{nik,gal02,qq1}, Chiral
\cite{chir3,ndim}, and many others
\cite{sa1,sa2,sa3,sa4,sa5,sa6,sa7,sa8,sa9,sa10,sa11}.

Brans-Dicke theory was introduced in the early sixties \cite{Brans} as a
gravitational theory which satisfies Mach's principle. In this gravitational
model, the scalar field is nonminimally coupled to gravity, which means that
the existence of the scalar field is essential for the gravitational field.
The theory is characterized by a parameter $\omega_{BD}$, known as the
Brans-Dicke parameter \cite{omegaBDGR}. Another major characteristic is that
the theory is defined in the Jordan frame \cite{Jord} and it is dynamically
equivalent to that of a minimally coupled scalar field in the Einstein frame
\cite{bb02,bb04}.

The physical properties of the gravitational model are not invariant under
conformal transformations; that is, the two conformal equivalent theories do
not share the same physical properties. We refer the reader to the discussion
in \cite{cc01}. A simple generalization of the Brans-Dicke model is to
introduce a non-constant parameter $\omega_{BD}=\omega_{BD}\left(
\phi\right)  $. This generalization leads to the so-called scalar-tensor
theories of gravity \cite{faraonibook}. Some applications of the scalar-tensor
theory for the description of physical phenomena can be found in
\cite{st1,st2,st3,st4,st5,st6,st7,st8,st9,st10,st11,st12} and references therein.

Within the framework of Bianchi Class A geometries, the scalar-tensor theories
admit a minisuperspace description \cite{mp1}. The existence of a
minisuperspace is essential for the quantization of the gravitational model
\cite{dw1,dw2,dw3} and the dynamical analysis of the gravitational field
equations, because the latter can be understood in the equivalent description
of point-like particles. Indeed, there exists a Lagrangian function for the
scalar-tensor theories which describes the motion of particles; as particles,
we consider the scale factors and the scalar field in a curved manifold under
the action of a conservative force, i.e. the scalar field potential. The
presence of the Lagrangian function allows us to apply various methods of
Classical Mechanics for the study of the field equations.

Noether's theorems \cite{ns0} have previously been used for the derivation of
variational symmetries and the construction of conservation laws for the
nonlinear gravitational field equations \cite{ns2,ns3,ns4}. The conservation
laws can be used for the reduction of the field equations by using invariant
functions and the calculation of exact and analytic solutions. There is a
plethora of applications in the literature on the Noether symmetry analysis in
gravity; for a review we refer the reader to \cite{ns1}. The Noether symmetry
analysis was applied for the study of scalar tensor cosmology in the case of
vacuum in \cite{ns4,lan1} and in the presence of a second scalar field in
\cite{nss1,nss2}.

In this work, we employ a different geometric method for the derivation of
analytic solutions, based on the Eisenhart-Duval lift \cite{el1,duval}. This
approach is the nonrelativistic limit of the Kaluza-Klein theory
\cite{kk1,kk2}, where the conservative forces are embedded into an extended
geometry by introducing new independent variables and degrees of freedom. In
this approach, the dynamical system is geometrized and is described by the
equivalent system of geodesic equations. Applications of the Eisenhart-Duval
method in classical and quantum field theory can be found in \cite{pap0,pap0a}%
, while some applications within the framework of cosmology can be found in
\cite{pap1,pap2,pap3,pap4,pap5,ppp1}.

The Eisenhart-Duval lift is a powerful method for demonstrating the geometric
linearization of dynamical systems. In \cite{pap6}, it was shown that
different gravitational models share the common solution, which is that of the
geodesic equations for flat space. The geometric linearization is performed
via the Eisenhart lift. Recently, in \cite{pap7}, new criteria and selection
rules were applied for the geometric linearization of constraint Hamiltonian
systems. Because in gravitational physics the point-like Lagrangian is
singular and describes a constraint Hamiltonian system, the geometric results
of \cite{pap7} can be applied. The structure of the paper is as follows.

In Section \ref{sec2}, we present the basic properties and definitions of
scalar-tensor theories. Emphasis is given to the minisuperspace description.
The Eisenhart-Duval lift for the scalar-tensor theory is presented in Section
\ref{sec3}. In Section \ref{sec4}, we present the geometric linearization of
the field equations and construct new analytic solutions for the field
equations. The Eisenhart-Duval lift in the case of a nonzero spatial curvature
universe is discussed in Section \ref{secn}. In Section \ref{secn2} we discuss
the application of conformal transformations to determine solutions for
conformal equivalent theories. Finally, in Section \ref{sec5}, we draw our conclusions.

\section{Scalar-Tensor Gravity}

\label{sec2}

We consider the generalized Brans-Dicke gravitational Action
\cite{faraonibook}%
\begin{equation}
S=\int dx^{4}\sqrt{-g}\left[  \frac{1}{2}\phi R-\frac{1}{2}\frac{\omega\left(
\phi\right)  }{\phi}g^{\mu\nu}\phi_{;\mu}\phi_{;\nu}-V\left(  \phi\right)
\right]  , \label{bd.01}%
\end{equation}
where $\omega\left(  \phi\right)  $ is a varying parameter and $V\left(
\phi\right)  $ is the scalar field potential. In the case where $\omega\left(
\phi\right)  =\omega_{BD}$ be a constant, the Brans-Dicke theory \cite{Brans}
is recovered.

Variation of the Action Integral (\ref{bd.01}) with respect to the metric
tensor leads to the gravitational field equations%
\begin{equation}
\phi G_{\mu\nu}=\frac{\omega\left(  \phi\right)  }{\phi^{2}}\left(  \phi
_{;\mu}\phi_{;\nu}-\frac{1}{2}g_{\mu\nu}g^{\kappa\lambda}\phi_{;\kappa}%
\phi_{;\lambda}\right)  -\frac{1}{\phi}\left(  g_{\mu\nu}g^{\kappa\lambda}%
\phi_{;\kappa\lambda}-\phi_{;\mu}\phi_{;\nu}\right)  -g_{\mu\nu}\frac{V\left(
\phi\right)  }{\phi}, \label{eq.01}%
\end{equation}
where $G_{\mu\nu}=R_{\mu\nu}-\frac{1}{2}Rg_{\mu\nu}$ is the Einstein tensor.

The field equations (\ref{eq.01}) can be written in the equivalent form%
\begin{equation}
G_{\mu\nu}=k_{eff}T_{\mu\nu}^{\phi},
\end{equation}
in which $k_{eff}=\frac{1}{\phi}$ is a varying effective gravitational
coupling parameter and $T_{\mu\nu}^{\phi}$ is the energy momentum tensor which
attributes the degrees of freedom related to the scalar field, that is,%
\begin{equation}
T_{\mu\nu}^{\phi}=\frac{\omega\left(  \phi\right)  }{\phi^{2}}\left(
\phi_{;\mu}\phi_{;\nu}-\frac{1}{2}g_{\mu\nu}g^{\kappa\lambda}\phi_{;\kappa
}\phi_{;\lambda}\right)  -\frac{1}{\phi}\left(  g_{\mu\nu}g^{\kappa\lambda
}\phi_{;\kappa\lambda}-\phi_{;\mu}\phi_{;\nu}\right)  -g_{\mu\nu}%
\frac{V\left(  \phi\right)  }{\phi}. \label{eq.02}%
\end{equation}

The equation of motion for the scalar field follows from the variation of
(\ref{eq.01}) with respect to the field $\phi$, that is,
\begin{equation}
g^{\kappa\lambda}\phi_{;\kappa\lambda}+\frac{1}{2}\left(  \ln\left(
\frac{\omega\left(  \phi\right)  }{\phi}\right)  \right)  _{,\phi}g^{\mu\nu
}\phi_{;\mu}\phi_{;\nu}+\frac{\phi}{2\omega\left(  \phi\right)  }\left(
R-2V_{,\phi}\right)  =0. \label{eq.03}%
\end{equation}

\subsection{FLRW\ Cosmology}

On very large scales, the universe is regarded as isotropic and homogeneous,
described by the spatially flat FLRW geometry, with the line element given by%

\begin{equation}
ds^{2}=-N^{2}\left(  t\right)  dt^{2}+a^{2}\left(  t\right)  \left(
dx^{2}+dy^{2}+dz^{2}\right)  , \label{bd.05}%
\end{equation}
in which $N\left(  t\right)  $ is the lapse function and $a\left(  t\right)  $
is the scale factor. For the comoving observer $u^{\mu}=\frac{1}{N}\delta
_{t}^{\mu}$, the expansion rate $\theta=u_{~;\mu}^{\mu}$is calculated
$\theta=3H^{2}$, with $H=\frac{1}{N}\frac{\dot{a}}{a}$ to be the Hubble
function. A dot represent total derivative with respect to the time parameter
$t$, that is, $\dot{a}=\frac{da}{dt}$. The line element (\ref{bd.05}) admits a
sixth-dimensional Killing algebra consisted by the isometries of the
three-dimensional flat space. The FLRW belongs to the family of Bianchi Class
A models. For the scalar field $\phi$, we assume that it inherits the
symmetries of the background geometry from where it follows $\phi=\phi\left(
t\right)  $, is only a function of the time parameter.

The Ricci scalar for the FLRW geometry (\ref{bd.05}) is calculated
\begin{equation}
R=\frac{6}{N^{2}}\left[  \frac{\ddot{a}}{a}+\left(  \frac{\dot{a}}{a}\right)
^{2}-\frac{\dot{a}\dot{N}}{aN}\right]  . \label{bd.06}%
\end{equation}
or equivalent in terms of the Hubble function%
\begin{equation}
R=6\left(  \frac{1}{N}\dot{H}+2H^{2}\right)  . \label{bd.06b}%
\end{equation}

We replace the expression for the Ricci scalar (\ref{bd.06}) in the Action
Integral (\ref{bd.01}) and integration by parts leads to the point-like
Lagrangian \cite{ns4}%
\begin{equation}
L\left(  N,a,\dot{a},\phi,\dot{\phi}\right)  =\frac{1}{2N}\left(  6a\phi
\dot{a}^{2}+6a^{2}\dot{a}\dot{\phi}+\frac{\omega\left(  \phi\right)  }{\phi
}a^{3}\dot{\phi}^{2}\right)  -a^{3}NV\left(  \phi\right)  .\label{eq.04}%
\end{equation}

Lagrangian function (\ref{eq.04}) describes a singular dynamical system with
Hamiltonian constraint given by the equation of motion $\frac{\partial
L}{\partial N}=0$. Indeed, variation with respect to the dynamical variables
leads to the equations of motion for the points%
\begin{align}
\frac{\partial L}{\partial N}  &  =0,\\
\frac{d}{dt}\left(  \frac{\partial L}{\partial\dot{a}}\right)  -\frac{\partial
L}{\partial a}  &  =0,\\
\frac{d}{dt}\left(  \frac{\partial L}{\partial\phi}\right)  -\frac{\partial
L}{\partial\phi}  &  =0.
\end{align}
Therefore, the gravitational field equations are%
\begin{equation}
3H^{2}+3H\frac{\dot{\phi}}{N\phi}+\frac{\omega\left(  \phi\right)  }{2}\left(
\frac{\dot{\phi}}{N\phi^{2}}\right)  ^{2}+\frac{V\left(  \phi\right)  }{\phi
}=0, \label{eq.05}%
\end{equation}%
\begin{equation}
\frac{2}{N}\dot{H}+3H^{2}+2H\frac{\dot{\phi}}{N\phi}-\frac{\omega\left(
\phi\right)  }{2}\left(  \frac{\dot{\phi}}{\phi N}\right)  ^{2}+\frac
{\ddot{\phi}}{N^{2}\phi}+\frac{V\left(  \phi\right)  }{\phi}-\frac{\dot{N}%
\dot{\phi}}{N^{3}\phi}=0,
\end{equation}%
\begin{equation}
\frac{\ddot{\phi}}{N^{2}\phi}+\left(  \ln\left(  \frac{\omega\left(
\phi\right)  }{\phi}\right)  \right)  _{,\phi}\frac{\dot{\phi}^{2}}{\phi
N^{2}}-\frac{\dot{N}\dot{\phi}}{N^{3}\phi}+3H\frac{\dot{\phi}}{N\phi}+\frac
{1}{\omega\left(  \phi\right)  }\left(  \left(  \frac{3}{N}\dot{H}%
+6H^{2}\right)  +V_{,\phi}\right)  =0.
\end{equation}

When the lapse function is a constant, i.e. $N\left(  t\right)  =1$, the
latter gravitational field equations are simplified as
\begin{equation}
3H^{2}+3H\frac{\dot{\phi}}{\phi}+\frac{\omega\left(  \phi\right)  }{2}\left(
\frac{\dot{\phi}}{\phi^{2}}\right)  ^{2}+\frac{V\left(  \phi\right)  }{\phi
}=0, \label{cs1}%
\end{equation}%
\begin{equation}
2\dot{H}+3H^{2}+2H\frac{\dot{\phi}}{\phi}-\frac{\omega\left(  \phi\right)
}{2}\left(  \frac{\dot{\phi}}{\phi}\right)  ^{2}+\frac{\ddot{\phi}}{\phi
}+\frac{V\left(  \phi\right)  }{\phi}=0,
\end{equation}%
\begin{equation}
\frac{\ddot{\phi}}{\phi}+\left(  \ln\left(  \frac{\omega\left(  \phi\right)
}{\phi}\right)  \right)  _{,\phi}\frac{\dot{\phi}^{2}}{\phi}+3H\frac{\dot
{\phi}}{\phi}+\frac{1}{\omega\left(  \phi\right)  }\left(  \left(  3\dot
{H}+6H^{2}\right)  +V_{,\phi}\right)  =0.
\end{equation}
When the lapse function has a specific function form, then we can say that the
Lagrangian (\ref{eq.04}) describes a regular dynamical system where the
constraint equation (\ref{cs1}) gives a specific value for the conservation
law of \textquotedblleft energy\textquotedblright\ for the point-like
equivalent dynamical system.

\section{Einsenhart-Duval lift in Scalar-Tensor Cosmology}

\label{sec3}

We introduce the new point-like Lagrangian function
\begin{equation}
\tilde{L}\left(  N,a,\dot{a},\phi,\dot{\phi},z,\dot{z}\right)  =\frac{1}%
{2N}\left(  6a\phi\dot{a}^{2}+6a^{2}\dot{a}\dot{\phi}+\frac{\omega\left(
\phi\right)  }{\phi}a^{3}\dot{\phi}^{2}+\frac{1}{a^{3}V\left(  \phi\right)
}\dot{z}^{2}\right)  \label{el.01}%
\end{equation}
defined in the extended space of variables $\left\{  a,\phi,z\right\}  $.
Scalar $z=z\left(  t\right)  $ has been introduced to attribute the
conservative forces as part of the geometry \cite{pap7}, such that the field
equations to be described as a set of geodesic equations.

Lagrangian function (\ref{el.01}) is singular, and it describes the (null-)
geodesic equations for the three-dimensional extended minisuperspace $g_{ij}$
with line element%
\begin{equation}
ds^{2}=6a\phi da^{2}+6a^{2}dad\phi+\frac{\omega\left(  \phi\right)  }{\phi
}a^{3}d\phi^{2}+\frac{1}{a^{3}V\left(  \phi\right)  }dz^{2}. \label{el.02}%
\end{equation}

The Hamiltonian constraint for the Lagrangian (\ref{el.01}) is%
\begin{equation}
3a\phi\dot{a}^{2}+3a^{2}\dot{a}\dot{\phi}+\frac{1}{2}\frac{\omega\left(
\phi\right)  }{\phi}a^{3}\dot{\phi}^{2}+\frac{1}{2a^{3}V\left(  \phi\right)
}\dot{z}^{2}=0, \label{el.03}%
\end{equation}
which indicate that the geodesic equations are null.

We observe that Lagrangian (\ref{el.01}) is independent of the scalar $z$,
thus, any transformation $z\rightarrow\bar{z}+z_{0}$, leaves Lagrangian
$\tilde{L}$ invariant. Consequently, the vector field $\partial_{z}$ is a
Noether symmetry with corresponding conservation law
\begin{equation}
I_{0}=\frac{\dot{z}}{a^{3}V\left(  \phi\right)  }. \label{el.04}%
\end{equation}
By replacing the latter conservation law in the constraint equation
(\ref{el.03}), the constraint equation (\ref{eq.05}) is recovered for
$I_{0}=\sqrt{2}$. Thus, Lagrangian function (\ref{el.01}) describes an
equivalent dynamical system with the scalar-tensor Lagrangian due to the
constraint $I_{0}=\sqrt{2}$. Mathematically, the two dynamical systems share
the same solution and the same algebraic properties.

To perform the global geometric linearization of the field equations, we
follow the approach described in \cite{pap7}.

Indeed, if the three-dimensional extended minisuperspace with line element
(\ref{el.02}) is conformally flat, then there exists a point transformation
$\left(  a,\phi,z\right)  \rightarrow\left(  X,Y,Z\right)  $, where the line
element (\ref{el.02}) becomes%
\begin{equation}
ds^{2}=M\left(  X,Y,Z\right)  \left(  \alpha_{1}dX^{2}+\alpha_{2}dY^{2}%
+\alpha_{3}dZ^{2}\right)  , \label{el.05}%
\end{equation}
and the null geodesics read%
\begin{equation}
\ddot{X}=0~,~\ddot{Y}=0~,~\ddot{Z}=0, \label{el.06}%
\end{equation}
with constraint equation%
\begin{equation}
\alpha_{1}\dot{X}^{2}+\alpha_{2}\dot{Y}^{2}+\alpha_{3}\dot{Z}^{2}=0.
\label{el.07}%
\end{equation}

Thus, the requirement for the three-dimensional space (\ref{el.02}) to be
conformally flat is that the Cotton-York tensor must have zero components. The
definition of the Cotton-York tensor is
\begin{equation}
C_{ijk}=R_{ij;k}^{g}-R_{kj;i}^{g}+\frac{1}{4}\left(  R_{;j}^{g}g_{ik}%
-R_{;k}^{g}g_{ij}\right)  ,
\end{equation}
in which $R_{ij}^{g}$ is the Ricci tensor and $R^{g}$ is Ricci scalar of the
extended minisuperspace $g_{ij}$.

It is of special interest to find a geometric gravitational Action Integral
which can produce the Lagrangian function (\ref{el.01}) such that to give a
physical meaning to the scalar $z\left(  t\right)  $. We introduce the
minisuperspace Lagrangian%
\begin{equation}
S=\int dx^{4}\sqrt{-g}\left[  \frac{1}{2}\phi R-\frac{1}{2}\frac{\omega\left(
\phi\right)  }{\phi}g^{\mu\nu}\phi_{;\mu}\phi_{;\nu}-\frac{1}{V\left(
\phi\right)  }F^{\mu\nu}F_{\mu\nu}\right]  , \label{el.07a}%
\end{equation}
where $F_{\mu\nu}=2A_{\left[  \mu,\nu\right]  }$ is the electromagnetic field
coupled to the scalar field. The coupling function is related to the potential
function $V\left(  \phi\right)  $ which drives the dynamics for the scalar
field. By introducing that $A_{\mu}=\int\frac{\dot{z}}{a^{2}}dt$, we end with
the point-like Lagrangian (\ref{el.01}).

\subsection{Geometric linearization}

Consequently, the geometric linearization criterion $C_{ijk}=0$, for the
extended minisuperspace (\ref{el.02}) leads to a set of nonlinear differential
equations which constrain the scalar field potential $V\left(  \phi\right)  $,
and coupling function $\omega\left(  \phi\right)  $.

It follows that the extended minisuperspace (\ref{el.02}) is conformally flat
for
\begin{equation}
V\left(  \phi\right)  =V_{0}\phi^{2}~,~\omega\left(  \phi\right)  \text{
arbitrary,} \label{el.08}%
\end{equation}
and when $V\left(  \phi\right)  ,~\omega\left(  \phi\right)  $ are related as
\begin{equation}
\omega\left(  \phi\right)  =\frac{3}{2}+\omega_{0}\left(  \left(  \phi\left(
\ln V\right)  _{,\phi}\right)  ^{2}+4\left(  1-4\left(  \ln V\right)  _{,\phi
}\right)  \right)  ~,~\omega\left(  \phi\right)  \text{ arbitrary.}%
\end{equation}
The latter gives the solutions
\begin{equation}
V\left(  \phi\right)  =V_{0}\phi^{2}\exp\left(  \pm\frac{2}{\sqrt{\omega_{0}}%
}\int\sqrt{\left(  \frac{\omega\left(  \phi\right)  }{\phi^{2}}-\frac{3}%
{2\phi^{2}}\right)  }d\phi\right)  . \label{el.09}%
\end{equation}

Therefore for the Brans-Dicke limit in which $\omega\left(  \phi\right)
=\omega_{BD}$, we find the two power-law potential functions%
\begin{equation}
V\left(  \phi\right)  =V_{0}\phi^{2\pm\frac{2}{\sqrt{\omega_{0}}}\sqrt{\left(
\omega_{BD}-\frac{3}{2}\right)  }}.
\end{equation}

We employ the following change of variables%
\begin{equation}
a=\alpha\phi\left(  \Phi\right)  ^{-\frac{1}{2}}~,~\Phi=\int\sqrt{\left(
\frac{\omega\left(  \phi\right)  }{\phi^{2}}-\frac{3}{2\phi^{2}}\right)
}d\phi, \label{el.10}%
\end{equation}
with%
\begin{equation}
\alpha=x^{A}y^{B}~,~e^{\frac{i}{\sqrt{6}}\Phi}=x^{-A}y^{B}, \label{el.11}%
\end{equation}
and $A=\frac{1}{\sqrt{6}\left(  \sqrt{6}+\lambda\right)  }~,~B=\frac{1}%
{\sqrt{6}\left(  \sqrt{6}-\lambda\right)  }$, $\lambda=\frac{2}{\sqrt
{-\omega_{0}}}$

In terms of the new coordinates the point-like Lagrangian function
(\ref{el.01}) for potential (\ref{el.09}) takes the canonical form of a
conformally flat space, that is,
\begin{equation}
L\left(  N,x,\dot{x},y,\dot{y}\right)  =\frac{1}{2N\phi\left(  \Phi\right)
^{\frac{1}{2}}}\left[  \frac{x^{-\frac{3i-\sqrt{6}\lambda}{6i-\sqrt{6}\lambda
}}y^{-\frac{3i+\sqrt{6}\lambda}{6i+\sqrt{6}\lambda}}}{V_{0}\left(
6+\lambda^{2}\right)  }\left(  4V_{0}\dot{x}\dot{y}+\left(  6+\lambda
^{2}\right)  \dot{z}^{2}\right)  \right]  .
\end{equation}
Therefore, the constraint equation reads%
\begin{equation}
\left(  4V_{0}\dot{x}\dot{y}+\left(  6+\lambda^{2}\right)  \dot{z}^{2}\right)
=0,
\end{equation}
and the gravitational field equations by applying the constraint equation
reads%
\begin{equation}
\ddot{x}=0~,~\ddot{y}=0~,~\ddot{z}=0. \label{el.12}%
\end{equation}

Hence, the analytic solution for this scalar-tensor cosmological theory is
constructed by the solution of the linear equations.

Consider now the case where $\omega\left(  \phi\right)  =\omega_{BD}$, then
from (\ref{el.10}) it follows $\Phi=i\sqrt{\frac{3}{2}-\omega_{BD}}\ln\phi$.
Thus, in the Brans-Dicke theory we calculate the analytic solution%
\begin{equation}
\alpha=x_{0}^{A}y_{0}^{B}\left(  t-t_{1}\right)  ^{A}\left(  t-t_{2}\right)
^{B},~\phi^{-}=x_{0}^{A\sqrt{\frac{3-2\omega_{BD}}{12}}}y_{0}^{-B\sqrt
{\frac{3-2\omega_{BD}}{12}}}\left(  t-t_{1}\right)  ^{A\sqrt{\frac
{3-2\omega_{BD}}{12}}}\left(  t-t_{2}\right)  ^{-B\sqrt{\frac{3-2\omega_{BD}%
}{12}}}.
\end{equation}
Thus, the scale factor reads%
\begin{equation}
a\simeq\left(  t-t_{1}\right)  ^{A\left(  1+\frac{1}{2}\sqrt{\frac
{3-2\omega_{BD}}{12}}\right)  }\left(  t-t_{2}\right)  ^{B\left(  1-\frac
{1}{2}\sqrt{\frac{3-2\omega_{BD}}{12}}\right)  },
\end{equation}
with Hubble function%
\begin{equation}
H=\frac{1}{N}\left(  \frac{A\left(  1+\frac{1}{2}\sqrt{\frac{3-2\omega_{BD}%
}{12}}\right)  }{\left(  t-t_{1}\right)  }+\frac{B\left(  1-\frac{1}{2}%
\sqrt{\frac{3-2\omega_{BD}}{12}}\right)  }{\left(  t-t_{2}\right)  }\right)  .
\end{equation}

\section{Modified theories of gravity}

\label{sec4}

One of the main characteristics of scalar-tensor theories is that the scalar
field can attribute the dynamical degrees of freedom introduced by the
modified theories of gravity, making the theories equivalent. We make use of
this equivalency in order to apply the results of the previous section and
derive analytic solutions for modified theories of gravity.

In particular, we study two modified theories of gravity, the $f\left(
R\right)  $-gravity \cite{fr1} and the hybrid metric-Palatini $f\left(
\mathcal{R}\right)  $-gravity~\cite{fr2}. These two theories have been widely
used for the description of cosmological and astrophysical phenomena, see for
instance \cite{fr3,fr4,fr5,fr6,fr7} and references therein.

\subsection{$f\left(  R\right)  $-cosmology}

In $f\left(  R\right)  $-theory, the gravitational Action Integral reads
\cite{fr1}%
\begin{equation}
S_{f\left(  R\right)  }=\int dx^{4}\sqrt{-g}f\left(  R\right)  \text{,}%
\end{equation}
where $f\left(  R\right)  $ is an arbitrary function of the Ricci scalar $R$
of the background space with metric $g_{\mu\nu}$. When $f\left(  R\right)  $
is a linear function the Einstein-Hilberrt Action is recovered and the theory
is reduced to that of General Relativity.

For the FLRW metric (\ref{bd.05}) and the Ricci scalar (\ref{bd.06}) we
introduce the Lagrange multiplier $\lambda_{m}$, such that the latter Action
Integral to be%
\begin{equation}
S_{f\left(  R\right)  }=\int dx^{4}\sqrt{-g}\left[  f\left(  R\right)
-\lambda\left(  R-\frac{6}{N^{2}}\left[  \frac{\ddot{a}}{a}+\left(  \frac
{\dot{a}}{a}\right)  ^{2}-\frac{\dot{a}\dot{N}}{aN}\right]  \right)  \right]
.
\end{equation}
Integration by parts leads to the following point-like Lagrangian%
\begin{equation}
L_{f\left(  R\right)  }\left(  N,a,\dot{a},R,\dot{R}\right)  =\frac{1}%
{N}\left(  6af_{,R}\dot{a}^{2}+6a^{2}f_{,RR}\dot{a}\dot{R}\right)
+Na^{3}\left(  f_{,R}R-f\right)  .
\end{equation}

We introduce the scalar field $\phi=f_{,R}$, and the function $V\left(
\phi\right)  =f-f_{,R}R$. Thus, Lagrangian $L_{f\left(  R\right)  }$ is
expressed in the equivalent form of the scalar-tensor theory, that is,
\begin{equation}
L_{f\left(  R\right)  }\left(  N,a,\dot{a},R,\dot{R}\right)  =\frac{1}%
{N}\left(  6a\phi\dot{a}^{2}+6a^{2}\dot{a}\dot{\phi}\right)  -Na^{3}V\left(
\phi\right)  . \label{lf.1}%
\end{equation}

By comparing (\ref{lf.1}) with (\ref{eq.04}) we observe that $\omega\left(
\phi\right)  =0$. Hence, the scalar field potential (\ref{el.09}) becomes%
\begin{equation}
V\left(  \phi\right)  =V_{0}\phi^{\mu}~,~\mu=2\pm\sqrt{6}\lambda
,~\lambda=\frac{2}{\sqrt{-\omega_{0}}},
\end{equation}
The power-law potential $V\left(  \phi\right)  =V_{0}\phi^{\mu}$ is related to
the power law theory $f\left(  R\right)  \simeq R^{\nu}$ with~$\mu=\frac
{\nu+1}{\nu}$.

Consequently, the transformation which linearize the field equations is
\begin{equation}
a=y^{B}~,~\phi=\left(  x^{-A}y^{B}\right)  ^{-\frac{\sqrt{6}}{3}},
\end{equation}
where $x,$~$y$ are given by the linear differential equations (\ref{el.12}).
At this point it is important to mention that in order for the solution to be
real, parameter $\omega_{0}$ must be negative.

Thus, the analytic solution for the scale factor reads%
\begin{equation}
a\left(  t\right)  =\left(  a_{0}\left(  t-t_{0}\right)  \right)  ^{B}\text{,}%
\end{equation}
and for the Hubble function we calculate%
\begin{equation}
H=\frac{1}{N}\frac{\dot{a}}{a}=\frac{1}{N}\frac{B}{t-t_{0}}.
\end{equation}

This solution was derived before by applying the Noether symmetry analysis in
\cite{anfr}. The algorithm of the Eisenhart-Duval lift is a simpler approach
for the construction of the analytic solution.

\subsection{Hybrid $f\left(  \mathcal{R}\right)  $-cosmology}

The gravitational Action Integral in hybrid metric-Palatini $f\left(
\mathcal{R}\right)  $-gravity is defined by the Ricci scalar $R$ for the
metric tensor and the Palatini curvature scalar $\mathcal{R}$ related to an
independent connection \cite{fr2}.

Specifically, the gravitational Action Integral is \cite{fr2}
\begin{equation}
S_{f\left(  \mathcal{R}\right)  }=\int d^{4}x\sqrt{-g}\left(  R+f(\mathcal{R}%
)\right)  , \label{s.01}%
\end{equation}

In a similar way with before, we introduce the scalar field $\psi
=f_{\mathcal{R}}$\bigskip$,$ and $V\left(  \psi\right)  =f\left(
\mathcal{R}\right)  -\mathcal{R}f^{\prime}\left(  \mathcal{R}\right)  $, and
with the use of the Lagrange multiplier we can write the latter Action
integral as%
\begin{equation}
S_{f\left(  \mathcal{R}\right)  }=\int d^{4}x\sqrt{-g}\left(  (1+\psi
)R+\frac{3}{2\psi}\partial^{\mu}\psi\partial_{\mu}\psi-V(\psi)\right)  .
\label{s.07}%
\end{equation}
We employ the change of variables $1+\psi=\phi$, thus the latter Action
Integral becomes%
\begin{equation}
S_{f\left(  \mathcal{R}\right)  }=\int d^{4}x\sqrt{-g}\left(  \phi R+\frac
{3}{2\left(  \phi-1\right)  }\partial^{\mu}\psi\partial_{\mu}\psi
-V(\phi)\right)  . \label{s.08}%
\end{equation}

In the case of a FLRW background geometry the minisuperspace Lagrangian reads
\begin{equation}
\left(  N,a,\dot{a},\phi,\dot{\phi}\right)  =\frac{1}{2N}\left(  6a\phi\dot
{a}^{2}+6a^{2}\dot{a}\dot{\phi}-\frac{3}{\left(  \phi-1\right)  }a^{3}%
\dot{\phi}^{2}\right)  -a^{3}NV\left(  \phi\right)  .
\end{equation}
Exact solutions for this gravitational model have been found before in
\cite{jr1,jr2}, while potential functions which lead to integrable field
equations were derived in \cite{jr3}.

Hence, the Action Integral (\ref{s.08}) is of the form of (\ref{bd.01}) with
\begin{equation}
\omega\left(  \phi\right)  =-\frac{3\phi}{\left(  \phi-1\right)  },
\end{equation}
and the corresponding potential function which linearize the dynamical system
is
\begin{equation}
V\left(  \phi\right)  =V_{0}\phi^{2}\exp\left(  \pm\frac{2}{\sqrt{\omega_{0}}%
}\int\sqrt{\left(  -\frac{3}{\phi\left(  \phi-1\right)  }-\frac{3}{2\phi^{2}%
}\right)  }d\phi\right)  ,
\end{equation}
that is,
\begin{equation}
V\left(  \phi\right)  =V_{0}\phi^{2}\left(  \frac{1+\frac{3\phi-1}%
{\sqrt{3\left(  \phi-1\right)  }}}{1-\frac{3\phi-1}{\sqrt{3\left(
\phi-1\right)  }}}\right)  ^{\frac{3\sqrt{2}}{\sqrt{-\omega_{0}}}}\left(
\frac{1-\sqrt{3}\phi+\sqrt{\left(  \phi-1\right)  \left(  3\phi-1\right)  }%
}{1+\sqrt{3}\phi-\sqrt{\left(  \phi-1\right)  \left(  3\phi-1\right)  }%
}\right)  ^{-\frac{\sqrt{6}}{\sqrt{-\omega_{0}}}}. \label{ff1}%
\end{equation}
and%
\begin{equation}
V\left(  \phi\right)  =V_{0}\phi^{2}.
\end{equation}
We remark that for large values of $\phi$ , potential (\ref{ff1}) has the
asymptotic limit $V\left(  \phi\right)  \simeq V_{0}\phi^{2}$.

Thus, the approach described above can be used for the derivation of the
analytic solution. These potential functions have not been derived before for
the case of hybrid metric-Palatini $f\left(  \mathcal{R}\right)  $-gravity.

\section{Nonzero spatial curvature}

\label{secn}

We proceed our analysis by considering a nonzero spatial curvature for the
FLRW line element, some integrable scalar tensor theories with spatial
curvature are presented in \cite{kkk1,kkk2}.

For this cosmological model, the point-like Lagrangian (\ref{eq.04}) is
modified as follows%
\begin{equation}
L_{K}\left(  N,a,\dot{a},\phi,\dot{\phi}\right)  =\frac{1}{2N}\left(
6a\phi\dot{a}^{2}+6a^{2}\dot{a}\dot{\phi}+\frac{\omega\left(  \phi\right)
}{\phi}a^{3}\dot{\phi}^{2}\right)  -a^{3}NV\left(  \phi\right)  +6Ka\phi,
\end{equation}
where $K$ is the spatial curvature.

We employ the Eisenhart-Duval lift and we introduce the extended
minisuperspace Lagrangian%
\begin{equation}
\tilde{L}_{K}\left(  N,a,\dot{a},\phi,\dot{\phi},z,\dot{z}\right)  =\frac
{1}{2N}\left(  6a\phi\dot{a}^{2}+6a^{2}\dot{a}\dot{\phi}+\frac{\omega\left(
\phi\right)  }{\phi}a^{3}\dot{\phi}^{2}+\frac{1}{\left(  a^{3}V\left(
\phi\right)  -6Ka\phi\right)  }\dot{z}^{2}\right)  \text{.}%
\end{equation}
Hence, the requirement the latter Lagrangian to describe a set of (null-)
geodesic equations for a conformally flat space leads to the constraint
equation%
\begin{equation}
\omega\left(  \phi\right)  =\frac{6\phi\left(  V^{2}\right)  _{,\phi}%
-3\phi^{2}V_{,\phi}^{2}}{2V^{2}},
\end{equation}
that is,
\begin{equation}
V\left(  \phi\right)  =V_{0}\phi^{2}\exp\left(  \pm\sqrt{\frac{2}{3}}\int
\sqrt{\left(  \frac{3}{2\phi^{2}}-\frac{\omega\left(  \phi\right)  }{\phi^{2}%
}\right)  }d\phi\right)  \text{.}%
\end{equation}
which is a specific case for potential (\ref{el.09}) with $\omega_{0}=-6$. \ 

\section{\textbf{Conformal transformations}}

\label{secn2}

It is well known that scalar tensor theories are related through conformal
transformations \cite{cc0,cc1,cc2,cc3,cc4,cc5}. Specifically, conformal
equivalent theories have common solution but different physical properties.
Observable quantities are not invariant under conformal transformations, for
more details we refer to the discussion in \cite{cc6,cc7}.

Let us demonstrate the conformal transformation in the case of the
minisuperspace description. Under the change of variables $N=\bar{N}%
\phi^{-\frac{1}{2}}$,~$a=s\phi^{-\frac{1}{2}}$, the point-like Lagrangian
(\ref{eq.04}) becomes
\begin{equation}
L\left(  \bar{N},s,\dot{s},\phi,\dot{\phi}\right)  =\frac{1}{\bar{N}}\left(
3s\dot{s}^{2}+\frac{1}{2}\left(  \frac{\omega\left(  \phi\right)  }{\phi^{2}%
}-\frac{3}{2\phi^{2}}\right)  a^{3}\dot{\phi}^{2}\right)  -a^{3}\bar{N}%
\frac{V\left(  \phi\right)  }{\phi^{2}}.
\end{equation}
By introducing the new scalar field $d\Phi=\sqrt{\left(  \frac{\omega\left(
\phi\right)  }{\phi^{2}}-\frac{3}{2\phi^{2}}\right)  }d\phi$ we end with a
minimally coupled scalar field, defined in the Einstein frame with effective
potential $\hat{V}\left(  \Psi\right)  =\frac{V\left(  \phi\right)  }{\phi
^{2}}$. Hence, the two potentials found before, that is, (\ref{el.08}), and
(\ref{el.09}) lead to the two effective potentials $\hat{V}\left(
\Psi\right)  =V_{0}$ and $\hat{V}\left(  \Psi\right)  =V_{0}\exp\left(
\pm\frac{2}{\sqrt{\omega_{0}}}\Psi\right)  $, that is, to the cosmological
constant and to the exponential potential.

On the other hand, under the conformal transformation $N=n\Psi\left(
\phi\right)  ^{\frac{1}{2}}\phi^{-\frac{1}{2}}$,~$a=b\Psi\left(  \phi\right)
^{\frac{1}{2}}\phi^{-\frac{1}{2}}$, in the point-like Lagrangian (\ref{eq.04})
we end with the Lagrangian function%
\begin{equation}
L\left(  n,b,\dot{b},\phi,\dot{\phi}\right)  =\frac{1}{2n}\left(  6\Psi\left(
\phi\right)  b\dot{b}^{2}+6a\dot{a}^{2}\dot{\Psi}+\frac{1}{2}a^{3}%
\Omega\left(  \phi\left(  \Psi\right)  \right)  \dot{\phi}^{2}\right)
-a^{3}n\frac{V\left(  \phi\right)  \Psi\left(  \phi\right)  }{\phi^{2}}%
\end{equation}
where $\Omega\left(  \phi\right)  =\frac{\left(  \left(  2\omega\left(
\phi\right)  -3\right)  F^{2}+3\phi^{2}F_{,\phi}^{2}\right)  }{2F^{3}\phi^{2}%
}$. Hence, by assuming $\Psi\left(  \phi\right)  $ to be the new scalar field
we end with the conformal equivalent theory%
\begin{equation}
L\left(  n,b,\dot{b},\phi,\dot{\phi}\right)  =\frac{1}{2n}\left(  6\Psi\left(
\phi\right)  b\dot{b}^{2}+6a\dot{a}^{2}\dot{\Psi}+\frac{1}{2}a^{3}\hat{\Omega
}\left(  \Psi\right)  \dot{\Psi}^{2}\right)  -a^{3}n\frac{V\left(  \phi\left(
\Psi\right)  \right)  \Psi}{\phi\left(  \Psi\right)  ^{2}}\text{.}%
\end{equation}
Consequently, the solutions derived before are related via conformal
transformations. That is, the solution for the hybrid metric-Palatini
$f\left(  \mathcal{R}\right)  $-gravity can be construct by the solution for
the $f\left(  R\right)  $-gravity, or the Brans-Dicke field via conformal transformations.

\section{Conclusions}

\label{sec5}

The derivation of exact and analytic solutions in gravitational physics is of
special interest. In modern cosmology, closed-form solutions provide us with
important information regarding the cosmological history and viability of the
given gravitational model. In light of this, in this study we investigate the
global geometric linearization for the field equations in scalar-tensor theory
within a spatially flat FLRW geometry. The cosmological field equations admit
a minisuperspace description, which means that techniques from analytic
mechanics can be applied.

In particular, in the previous Sections we employed the Einsenhart-Duval lift
as a means to write the gravitational field equations in the equivalent form
of geodesic equations for an extended minisuperspace. In this approach, the
potential terms which drive the cosmological dynamics are attributed to the
coupling function of a new scalar field.

The requirement for the extended minisuperspace to be conformally flat is
equivalent to the existence of a point transformation where the field
equations can be written as the equations of the free particle in a flat
space. The minisuperspace in scalar-tensor cosmology has dimension two, and
the extended minisuperspace has dimension three. Hence, the demand for the
component of the Cotton-York tensor of the extended minisuperspace to be
always zero, that is, the space must be conformally flat, leads to a system of
differential equations which constrain the free parameters of the theory.

The solution of the latter system provided us with two sets of values for the
scalar field potential and the coupling function.

For these specific sets of free functions, we provided the closed-form
expression for the point transformation which leads to the geometric
linearization of the field equations. As a result, the field equations have a
common solution expressed by the equations for the free particle in a flat
space, but in different coordinate systems. These results are applied for the
construction of analytic solutions in modified theories of gravity which are
equivalent to the scalar-tensor theory.

The Eisenhart-Duval lift is a powerful method for the study of the
integrability properties and the derivation of solutions in modern cosmology.
The Eisenhart-Duval lift is a complementary method to the variational
symmetries applied previously. The novelty of the Eisenhart-Duval lift is
mainly the geometric approach of the method and the use of simple geometric
criteria. Unlike in the Noether symmetry analysis, where the generator of the
point transformation should be defined and the Noether symmetry conditions
need to be solved, in this consideration of the Eisenhart-Duval lift we
require the extended minisuperspace to be conformally flat.

The definition of the lift, that is, of the extended minisuperspace, is not
unique, and the requirement for the space to be conformally flat can be
relaxed further to require that the extended minisuperspace is integrable.
This can lead to the derivation of new integrable cosmological models.

\textbf{Data Availability Statements:} Data sharing is not applicable to this
article as no datasets were generated or analyzed during the current study.

\begin{acknowledgments}
The author thanks the support of Vicerrector\'{\i}a de Investigaci\'{o}n y
Desarrollo Tecnol\'{o}gico (Vridt) at Universidad Cat\'{o}lica del Norte
through N\'{u}cleo de Investigaci\'{o}n Geometr\'{\i}a Diferencial y
Aplicaciones, Resoluci\'{o}n Vridt No - 098/2022.
\end{acknowledgments}

\end{document}